\documentclass[preprint,12pt]{elsarticle}

\usepackage{amssymb}

\biboptions{square,sort&compress}

\usepackage[footnotesize]{caption2}

\usepackage{url}

\journal{Physica D}

\begin{document}

\begin{frontmatter}



\title{An Empirical Investigation of Scaling Behavior in the Atmospheric Turbulence for Understanding the Underlying Cascade Process}


\author{Lei Liu}
\ead{liulei@mail.iap.ac.cn}
\author{Fei Hu\corref{cor1}}
\ead{hufei@mail.iap.ac.cn} \cortext[cor1]{Corresponding author}

\address{State Key Laboratory of Atmospheric Boundary Layer Physics and Atmospheric
Chemistry, Institute of Atmospheric Physics, Chinese Academy of
Sciences, Beijing, 100029 China}

\begin{abstract}
We study the scaling behaviors in the wind velocity time series
collected at the atmospheric surface layer and compare them with two
commonly used cascade models, the truncated stable distribution and
the log-normal model. Results show that although both models can
describe the change of probability density functions from
non-Gaussian to Gaussian like distributions with the increase of
time scale, they can not fit the scaling behaviors observed in the
probability of return and in the moments at the same time. This work
provides some clues on the understanding of cascade process in the
atmospheric turbulence.
\end{abstract}


\end{frontmatter}


\section{Introduction}
\label{intro}

The probability density functions (PDFs) of the atmospheric
turbulent velocity increments have been found to go from Gaussian
like behavior at a time scale of few days to stretched exponential
like behavior at a time scale of one hour \cite{mbp09}. This
behavior is similar to that observed in fully developed homogeneous
and isotropic turbulence, where it has long been recognized that the
energy is transferred from large scales to smaller scales by some
cascade process. Although the atmospheric turbulence is obviously
non-homogenous and non-isotropic, its statistical similarity with
homogeneous and isotropic turbulence suggests that there may be also
some cascade process associated with the energy transfer between
synoptic scales and smaller scales in the atmospheric turbulence.

To our knowledge, two models are commonly used to describe the PDFs
of atmospheric turbulent velocity increments. One is the log-normal
model which was proposed to account for the intermittency in the local homogeneous and isotropic turbulence \cite{cgh91}. The other is the
truncated stable distribution (sometimes called truncated L\'{e}vy
distribution), which was originally proposed to resolve the paradox
between infinite variance of stable distribution and finite variance
of real economic systems \cite{ms94,ms95}.
Some researches show that both models are also good enough to fit the PDFs of velocity in atmospheric turbulence \cite{brwp03,liu08,liu11} . However, as we will see in this
study, the underlying cascade processes of the two models are radically different while they are mathematically equivalent for data fitting.

In addition to PDFs, some other functions can also be used to reflect the underlying cascade process from the aspect of scaling. In homogeneous and isotropic turbulence, the moments of turbulent velocity increments between two points separated by a time or spatial interval (called time or spatial scale in this paper) are always used as a tool to reveal the underlying cascade process. Generally, the $q$th-order moments vary as a power function of scales and their power exponents are concave with respect of $q$. An another function usually discussed in the stable or truncated stable distributions is the probability of return, which are also found to be a power of scales when scales are small. In fact, this scaling behavior reflects a self-affine cascade process in the stable or truncated stable distributions at small scales\cite{ms95, mandelbrot97}.

The aim of this paper is to analyze the scaling behaviors (i.e., the probability of return and the moments as a power of time scales) in the time series of atmospheric turbulent velocity increments, by using a large amount of experimental data collected in the atmospheric boundary layer (see Sec.~\ref{scal}). Then, the results are compared with above mentioned PDF models each with a special cascade process. The inconsistencies between these models and observed scaling behaviors are discussed in detail (see Sec.~\ref{comp}). It hopes that our research will give some clues about cascade process in the atmospheric turbulence.

\section{PDFs of velocity increments}
\label{pdfs}

Before studying scaling behaviors possibly existing in the atmospheric turbulent velocity
increments, we first analyze their probability density functions (PDFs) and compare the results with models. Data used in this study is collected at a site located in a steppe of northeast Xilinhaote city, in Inner Mongolia, China, where wind velocities are measured at a height of 30 m by means of a sonic anemometer with a sampling frequency of 20 Hz (Campbell
CSAT-3). A velocity time series of continuous 320 hours observations (i.e., from 13 October 2009 until 26 October 2009) is chosen to analyze PDFs. Thus, as many as $2\times 10^7$ samples are used here, which can give a better estimate of PDFs especially their tails.

In the sequel, $v(t)$ will denote the modulus of a wind velocity
vector while it is used to analyzing the dimensionless value
$\hat{v}(t)$:
\begin{equation}
\hat{v}(t)=\frac{v(t)}{\sigma_{v}}, \label{eq:1}
\end{equation}
where $\sigma_{v}$ is the standard deviation of $v(t)$. The
increments of normalized wind velocity between two points separated by a time interval of $\tau$ (i.e, by a time scale of $\tau$) are then defined by
\begin{equation}
\triangle \hat{v}_{\tau}=\hat{v}(t+\tau)-\hat{v}(t).
\label{eq:2}
\end{equation}
In fact, the non-overlapping velocity increments can also be estimated by the sum of velocity increments at smallest time scale:
\begin{equation}
\triangle \hat
{v}_{\tau}=\sum_{i=1}^{\tau/\tau_0}[\hat{v}(t+i\tau_0)-\hat{v}(t+(i-1)\tau_0)].
\label{eq:3}
\end{equation}
where $\tau_0$ is the smallest time scale and equals to the
reciprocal of sampling frequency.

\begin{figure}
\centering
 \noindent\includegraphics[width=16pc]{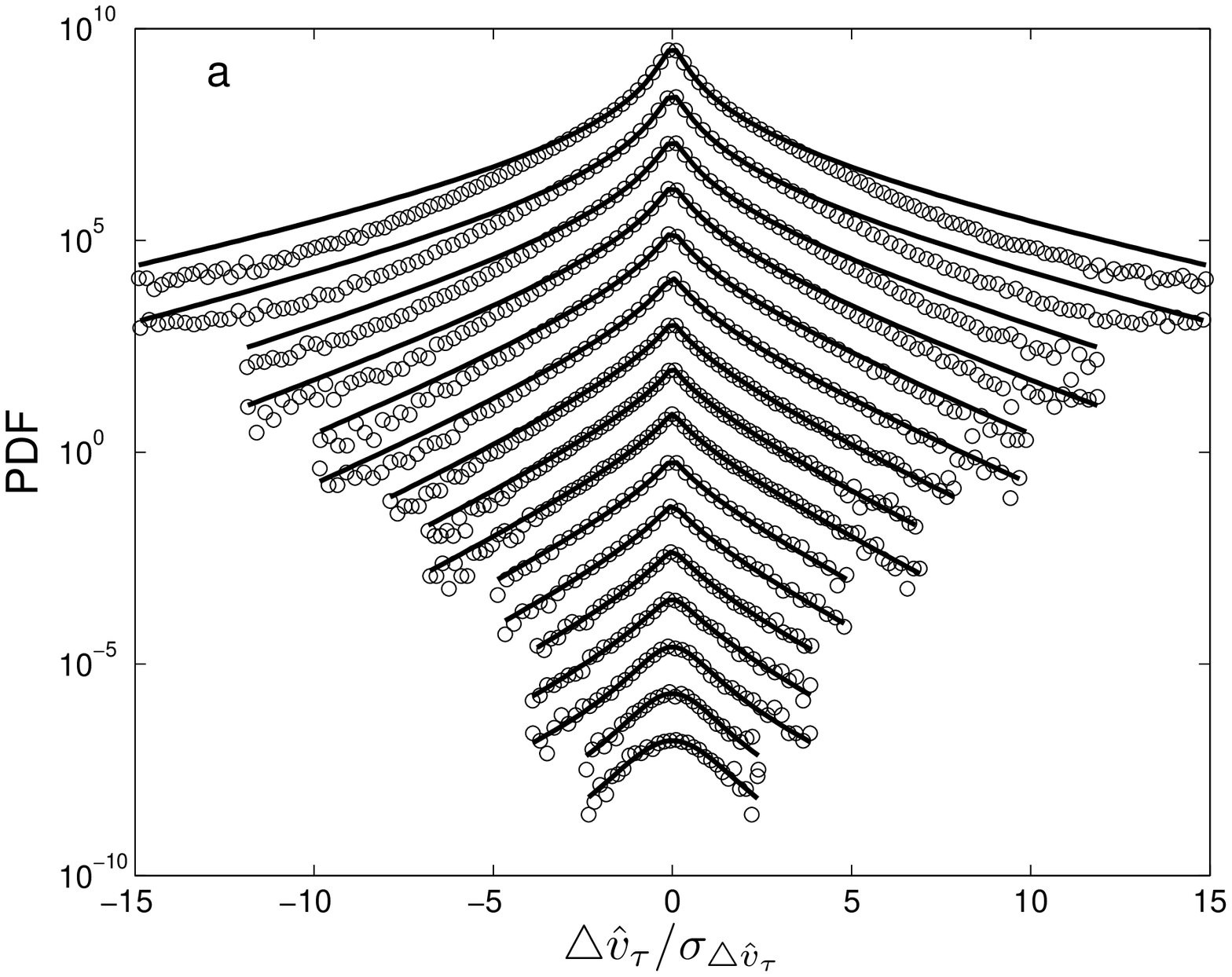}
 \noindent\includegraphics[width=16pc]{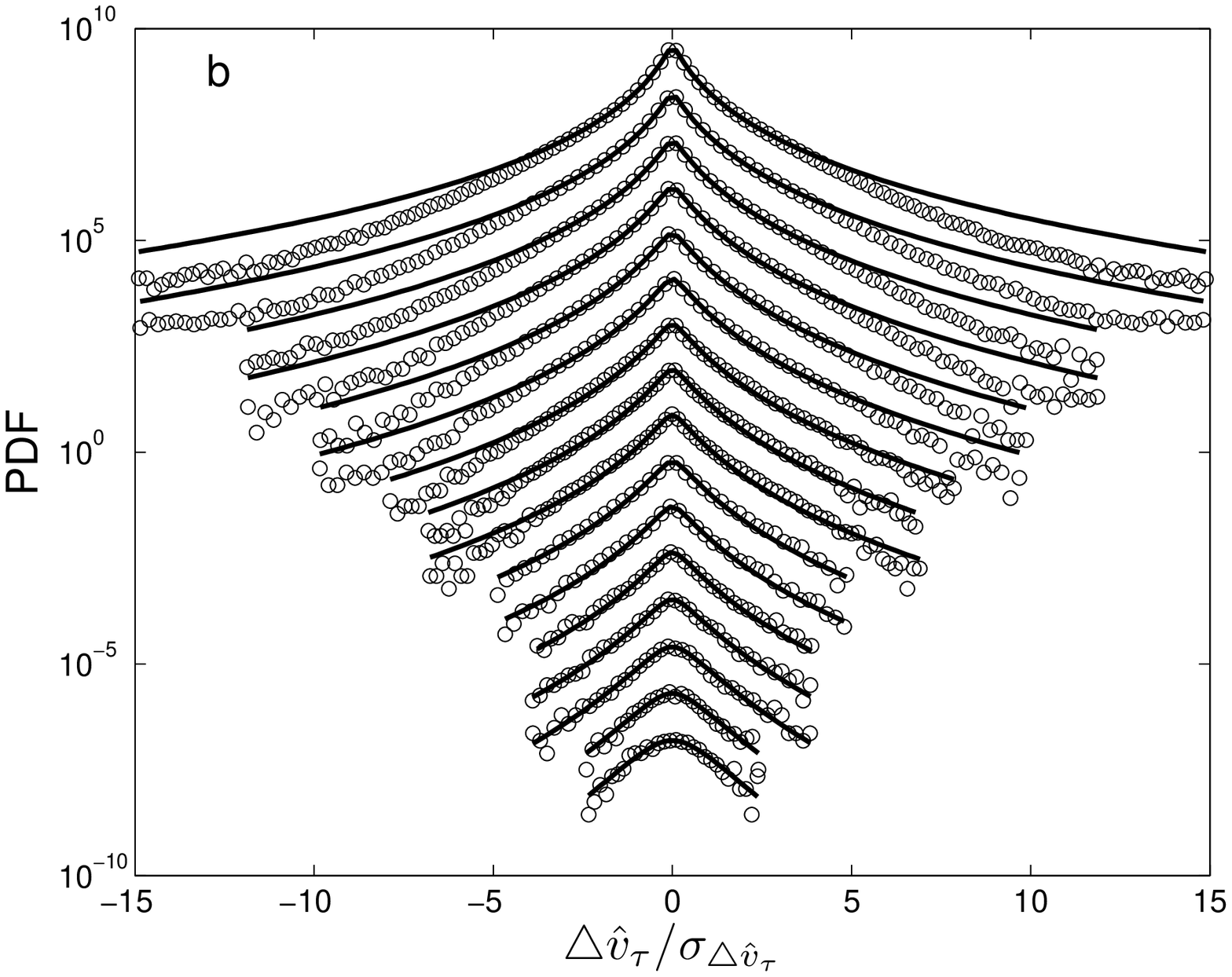}
 \caption{\label{fig:1} Comparison of standardized PDFs of velocity increments
 $\triangle \hat{v}_{\tau}$ at different time scales (circles) with
 (a) the truncated stable distributions (lines) and (b) log-normal PDF model
(lines). From top to bottom, time scales which range from $0.05$ s
to $1638.4$ s increase as a geometric sequence with a common ratio
of 2. The symbols $\sigma_{\triangle \hat{v}_{\tau}}$ is the
standard deviation of $\triangle \hat{v}_{\tau}$. Plots have been
arbitrarily shifted for illustration. }
 \end{figure}

Standardized PDFs of $\triangle \hat{v}_{\tau}$ at different time
scales are shown in Fig.~\ref{fig:1}. One can see that these PDFs
are almost symmetrical and go from Gaussian like behavior at a time
scale of more than ten minutes to some non-Gaussian behavior (with
longer tails than Gaussian) at smaller time scales. This result is
similar to that in Ref.~\cite{mbp09} (see Fig.~4a in this article)
where a longer time series (from 1960 to 1999) but with lower
sampling frequency (1 hour) collected at the atmospheric boundary
layer has been used. Moreover, this deformation behaviors of PDFs in
the atmospheric turbulence are also similar to those observed in the
local homogeneous and isotropic turbulence (see Fig.~4b in
Ref.~\cite{mbp09}). In the latter, it has long been recognized that
the energy is transferred between various scales by the cascade
process. It suggests that the energy transfer between atmospheric
mesoscales and more smaller scales may be also related to some
cascade process, although the turbulence at these scales is
non-homogeneous and isotropic and is influenced by many factors like
frictional drag, evaporation and transpiration, heat transfer,
pollutant emission, terrain induced flow modification and so on
\cite{stull}.

In this study, two commonly used PDF models will be screened. One is the truncated stable distribution and the other is the log-normal PDF model. The truncated stable
distribution was first proposed in Ref.~\cite{ms94} and then was used to
account for the scaling behavior in the dynamics of an economic index
\cite{ms95}. After that, many other truncated stable distributions have been proposed
\cite{koponen,gc99,gc00,mrd03}. The distinction between these distributions is mainly focus on their tail distributions. In this research, we will focus on the one
proposed in Ref.~{\cite{koponen}}, which has the fewest tuning
parameters and a smooth form at the same time. In
Refs.~\cite{liu08,liu11}, this truncated stable distribution has been
reported to fit the PDFs of atmospheric turbulent velocity. This distribution dose not
have an analytical expression but has an analytical characteristic function
$\Phi(k)$ \cite{koponen}:
\begin{eqnarray}
\ln \Phi(k)&=&\displaystyle
\frac{\gamma^{\alpha}}{\cos(\pi\alpha/2)}\left\{\lambda^\alpha-
(k^2+\lambda^2)^{\alpha/2}\cos
\left(\alpha\arctan\frac{k}{\lambda}\right)\right.\nonumber\\
&&\displaystyle \left.\left[1-i\beta\tan\left(\alpha
\arctan\frac{k}{\lambda}\right)\right]\right\}, \label{eq:4}
\end{eqnarray}
when $0<\alpha<1$, and
\begin{eqnarray}
\ln\Phi(k)&=&\displaystyle
\frac{\gamma^{\alpha}}{\cos(\pi\alpha/2)}\left\{\lambda^\alpha-
(k^2+\lambda^2)^{\alpha/2}\cos\left(\alpha\arctan\frac{k}{\lambda}\right)\right.\nonumber\\
&&\displaystyle \left[1-i\beta\tan\left(\alpha
\arctan\frac{k}{\lambda}\right)\right]\nonumber\\
&&-i\alpha\beta\lambda^{\alpha-1}k\bigg\}, \label{eq:5}
\end{eqnarray}
when $1<\alpha<2$. Symbols $\alpha$, $\beta$, $\gamma$ and
$\lambda$ represent four independent parameters which are normally
obtained by fitting Eqs.~(\ref{eq:4}) and (\ref{eq:5}) with the
data. Equations (\ref{eq:4}) and
(\ref{eq:5}) can be written in a simpler form \cite{nakao}:
\begin{equation}
\ln
\Phi(k)=C\Gamma(-\alpha)[q(\lambda+ik)^\alpha+p(\lambda-ik)^\alpha-\lambda^\alpha],
\label{eq:6}
\end{equation}
when $0<\alpha<1$, and
\begin{equation}
\ln\Phi(k)=C\Gamma(-\alpha)[q(\lambda+ik)^\alpha+p(\lambda-ik)^\alpha-\lambda^\alpha-i\alpha\lambda^{\alpha-1}(q-p)k],
\label{eq:7}
\end{equation}
when $1<\alpha<2$. The parameters are
$C=-\gamma^\alpha/[\cos(\alpha\pi/2)\Gamma(-\alpha)]$,
$p=(1+\beta)/2$ and $q=(1-\beta)/2$. For standardized and
symmetrical PDFs, $\beta=0$ and \cite{liu11}:
\begin{equation}
C\Gamma(-\alpha)\alpha\lambda^{\alpha-2}(\alpha-1)=1. \label{eq:8}
\end{equation}

The log-normal model was originally proposed to account for the PDFs
of velocity increments in the fully developed homogeneous and
isotropic turbulence \cite{cgh91}. This model does not have an
analytical PDF either. Its symmetrical PDF is:
\begin{equation}
f_\tau(\triangle
\hat{v})=\frac{1}{2\pi\lambda_{\tau}}\int_{0}^{\infty}\exp\left(-\frac{\triangle
\hat{v}^2}{2\sigma^2}\right)\exp\left(-\frac{\ln^2(\sigma/\bar{\sigma}_{\tau})}{2\lambda_{\tau}^2}\right)\frac{d\sigma}{\sigma^2},
\label{eq:9}
\end{equation}
where $\bar{\sigma}_{\tau}$ and $\lambda_{\tau}$ are parameters depending on scale $\tau$. For standardized PDF, there is
only one independent parameter and the other parameter can be obtained by
resolving the equation:
\begin{equation}
\bar{\sigma}_{\tau}^2\exp(2\lambda_{\tau}^2)=1. \label{eq:17}
\end{equation}

Figure \ref{fig:1} shows the comparison of experimental PDFs with
above two models. Results show that the truncated stable distribution
seems as ``good'' as the log-normal model when fitting to the data.
They both fit to the experimental PDFs well except their far tails where the
data seem to decay faster than the models with unknown reasons (see also
Fig.~2 in Ref.~\cite{brwp03}, where a similar deviation from
log-normal model has been observed). This reason may be
physical, but also may be artificial. Due to the limited data, tail distributions may be underestimated. In short, by only
fitting the PDFs we cannot say which model is better to account for
the data. We also cannot say anything more about the underlying cascade
process in the atmospheric turbulence. This is the motivation for us to look
into the scaling behaviors in the atmospheric turbulence.

\section{Scaling behavior}
\label{scal}

\subsection{Probability of return}
\label{prob}

\begin{figure}
\centering
 \includegraphics[width=20pc]{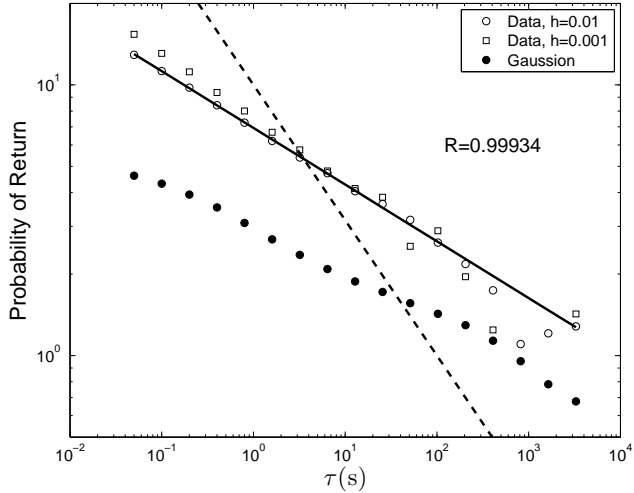}
 \caption{\label{fig:2} Probability of return for the
 wind velocity increments as a function of time scales $\tau$ (circles
 for threshold $h=0.01$ and squares for $h=0.001$). For comparison,
 the probability of return for Gaussian
 distributions (see Eq.~(\ref{eq:19})) is also shown in this plot (black dots). The
 slope of the best-fit line (fitting data when $\tau<204.8$ s) is -0.2145 and
 the corresponding correlation coefficient is about 0.999. Broken line shows a power function with
 an exponent of -0.5.}
 \end{figure}

Probability of return is defined as the probability density at the origin for symmetrical probability density function (PDF). In the data analysis, a probability density defined on a small interval $(-h, h)$ can be used as an approximation of the probability of return:
\begin{equation}
f_{\tau}(\triangle \hat{v}=0)\approx f_{\tau}(-h<\triangle
\hat{v}<h), \label{eq:11}
\end{equation}
if the threshold $h$ is small enough. In Fig.~\ref{fig:2}, circles denote the probability of return of atmospheric turbulent velocity with a threshold of 0.01. At small scales, this function behaves
as a power of time scales $\tau$:
\begin{equation}
f_{\tau}(\triangle \hat{v}=0)\propto \tau^{-p}. \label{eq:18}
\end{equation}
A least-square fitting shows that the exponent $p\approx 0.2145$.
For comparison, the probability of return of Gaussian distribution
is also shown here:
\begin{equation}
G_\tau(0)=\frac{1}{\sqrt{2\pi}\sigma_\tau}, \label{eq:19}
\end{equation}
where the standard deviation $\sigma_\tau=\sigma_{\triangle
\hat{v}_\tau}$. The difference between $f_{\tau}(\triangle
\hat{v}=0)$ and $G_\tau(0)$ decreases with the increase of $\tau$,
which implies a convergence to the gaussian distribution at large time scales. We also compute the probability of return with a much
smaller threshold of 0.001 and find that above results do not
change any more except that there are some statistical errors at large time scales and a little larger slope at small time scales.

\subsection{Moments}
\label{mome}

\begin{figure}
\centering
 \noindent\includegraphics[width=16pc]{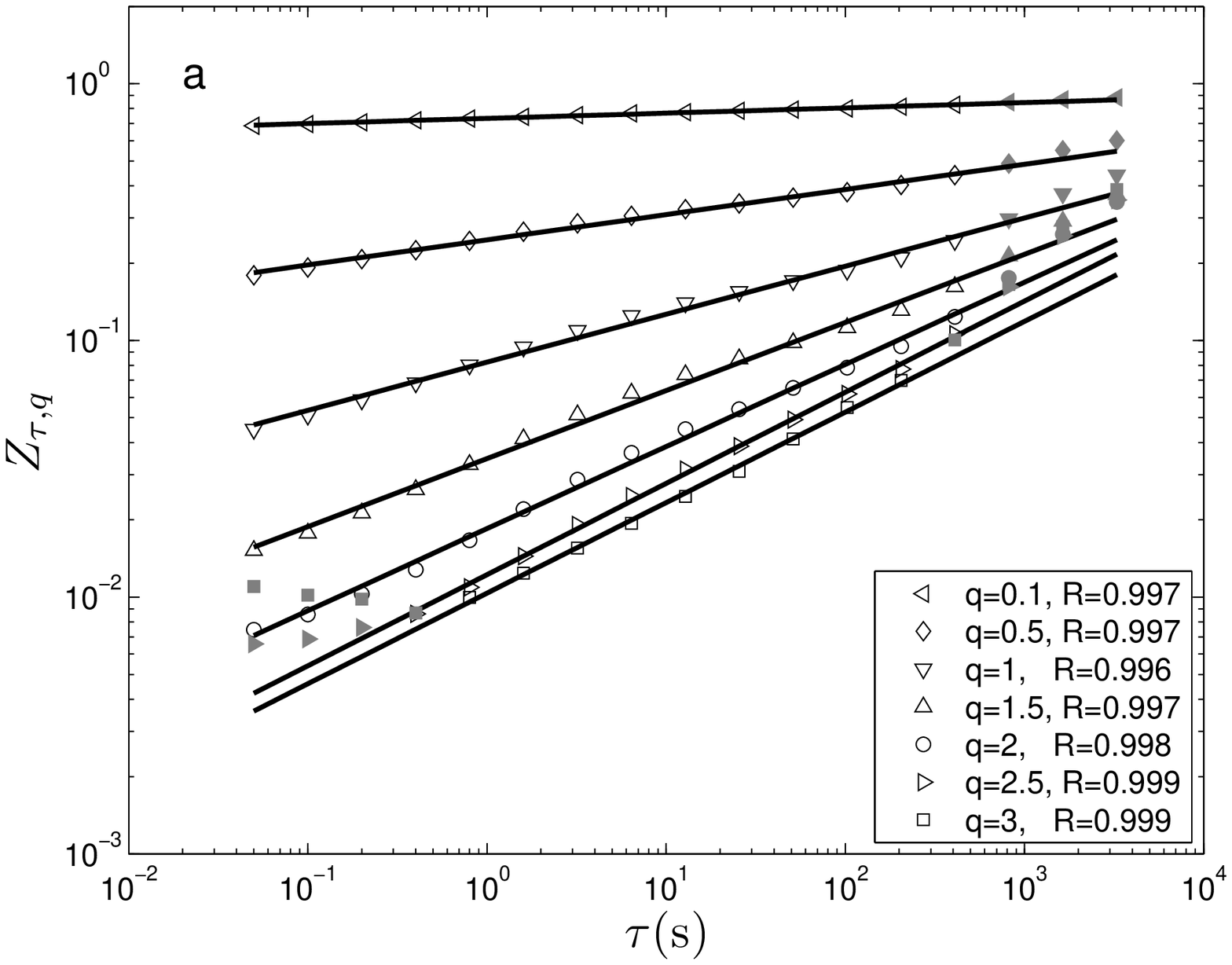}
 \noindent\includegraphics[width=16pc]{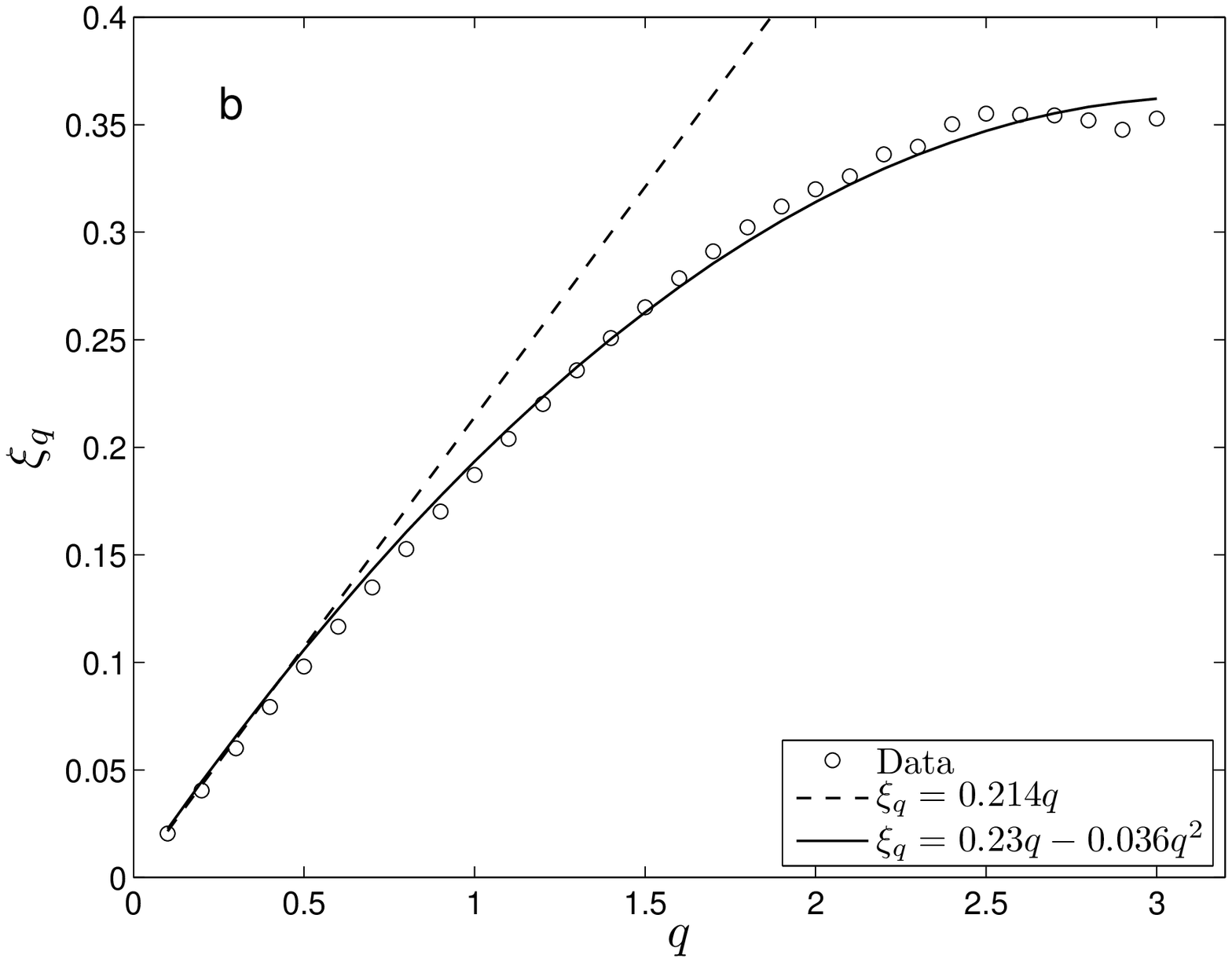}
 \caption{\label{fig:3} (a) Scaling behavior of $q$th order moments $Z_{\tau,q}$. Lines represent the least-square
 fitting to Eq.~(\ref{eq:13}) and the corresponding correlation coefficients $R$
 are also listed. Grey points are not used
 in the fittings. (b) Scaling exponents $\xi_q$ vary as a function of $q$.
 Line shows the fitting result with the log-normal model and
 the broken one is a line through the origin with a slope of $p$,
 where $p$ is the scaling exponent of probability of return (see Eq.~(\ref{eq:18})).}
 \end{figure}

Scaling behavior in the statistical moments of velocity increments (also called structure
functions) is the central topic for the homogeneous and isotropic turbulence. Many theoretical works have devoted to the understanding of this scaling behavior
and the underlying physical mechanisms \cite{frisch}. For
the non-homogeneous and non-isotropic atmospheric turbulence, its
statistical moments also have scaling behavior at a wide range
of time scales. Figure \ref{fig:3}a shows the statistical moments of
atmospheric turbulent velocity increments. One can see that at small
time scales,
\begin{equation}
Z_{\tau,q}\equiv \overline{|\triangle \hat{v}_{\tau}|^q}\propto
\tau^{\xi_q},
\label{eq:13}
\end{equation}
where the exponents $\xi_q$ vary as a function of $q$
(Fig.~\ref{fig:3}b). This power-law scaling behavior is observed for
time intervals spanning three orders of magnitude for small values
of $q$. However, for larger values of $q$ (for example, $q=2.5$ and
$3$ in Fig.~\ref{fig:3}a) data begin to deviate from this scaling
behavior at very small time scales ($\tau<1s$). The higher-order
moments is mainly defined by the tail distributions which describe
the statistical behavior of turbulent eddies with large
characteristic velocities. It suggests that the relatively very
small turbulent eddies with large characteristic velocities may have
different cascade process from these eddies with larger scales.
However, as already remarked in Fig.~\ref{fig:1}, all the
conclusions about the tail distributions should be carefully
screened because of the limited data.

Comparing with the exponents $\xi_q$ of homogeneous and isotropic
turbulence \cite{frisch}, we found that $\xi_q$ of non-homogeneous
and non-isotropic atmospheric turbulence are also concave but their
values are much smaller. For the homogeneous and isotropic
turbulence at high Reynolds number, the slope of $\xi_q$ at small
values of $q$ is about $1/3$ and $\xi_3=1$ (Kolmogorov's four-fifths
law). As shown in Fig.~\ref{fig:3}b, the slope of $\xi_q$ at small
values of $q$ in the atmospheric turbulence  is about $1/5$ and
$\xi_3\approx 0.35$. According to the Kolmogorov hypothesis in the
log-normal model\cite{kolmogorov62,cgh91},
\begin{equation}
9\lambda^2=\Lambda^2_{0}-\mu\ln \tau, \label{eq:12}
\end{equation}
where $\Lambda_0$ and $\mu$ are parameters independent of $\tau$.
For a fixed time scale, a smaller value of $\mu$ corresponds to
a larger value of $\lambda$ which means that the tail of PDF becomes longer
and the corresponding turbulence is more intermittent (see Eq.~(\ref{eq:9})). Following Eq.~(\ref{eq:12}) and
the assumption $\bar{\sigma}_\tau\propto\tau^b$, one can deduce that the exponents $\xi_q$ behave as a simple parabola:
\begin{equation}
\xi_q=bq-aq^2,
\label{eq:14}
\end{equation}
where $b$ is some constant and $a=\mu/18$. A least-square fitting shows that $a\approx 0.04$ for atmospheric turbulence while
this value is about 0.01 observed in the fully developed homogeneous and
isotropic turbulence \cite{frisch}. Thus, from the view of log-normal
model and the Kolmogorov hypothesis the homogeneous and isotropic
turbulence is less intermittent than the non-homogeneous and
non-isotropic atmospheric turbulence.

\section{Comparison with models}
\label{comp}

The truncated stable distribution and the log-normal model have
different cascade processes which can cause the different scaling
behaviors. The center part of truncated stable distribution at small
scales behaves like a stable distribution with the same parameters
$\alpha$, $\beta$ and $\gamma$. A notable feature of the stable distribution is its stability \cite{nolan}:
\begin{equation}
\sum_{i=1}^{n}\triangle
\hat{v}_{\tau_0,i}\stackrel{d}{=}n^{\frac{1}{\alpha}}\triangle
\hat{v}_{\tau_0}, \label{eq:15}
\end{equation}
where $n$ is a positive integer and $\triangle
\hat{v}_{\tau_0,i}$ $(i=1,2,\cdots,n)$ are independent random
variables with the same distribution as $\triangle
\hat{v}_{\tau_0}$. The symbol ``$\stackrel{d}{=}$'' means equality in
distribution.

From Eq.~(\ref{eq:3}), one can see that
\begin{equation}
\triangle \hat {v}_{\tau}=\sum_{i=1}^{n}\triangle \hat
{v}_{\tau_0,i}, \label{eq:10}
\end{equation}
where $\triangle \hat
{v}_{\tau_0,i}\equiv \hat{v}(t+i\tau_0)-\hat{v}(t+(i-1)\tau_0)$ and $n=\tau/\tau_0$. Thus, the truncated stable distribution contains a self-affine
cascade process at small scales \cite{mandelbrot97}:
\begin{equation}
\triangle \hat
{v}_{\tau}\stackrel{d}{=}n^{\frac{1}{\alpha}}\triangle
\hat{v}_{\tau_0}\label{eq:20}.
\end{equation}
An equivalent expression of Eq.~(\ref{eq:20})
is
\begin{equation}
f_{\tau_0}(\triangle
\hat{v})=n^{1/\alpha}f_{\tau}(n^{1/\alpha}\triangle \hat{v}).
\label{eq:16}
\end{equation}
From this expression, one can deduce that the probability
of return for the truncated stable distribution behaves as a power of time scale:
\begin{equation}
f_{\tau}(\triangle \hat{v}=0)\propto \tau^{-1/\alpha}. \label{eq:21}
\end{equation}
Although the probability of return of truncated stable distribution has a scaling behavior at small scales, one should note that the scaling exponent
$1/\alpha\in [0.5,\infty)$. This scaling exponent is larger than that observed in the atmospheric turbulence (see Fig.~\ref{fig:2}). For the $q$th-order moments, authors
(Refs.~\cite{nakao,twp06}) have proved that the truncated stable
distribution also has a scaling behavior but the scaling exponents vary as a bi-linear function of $q$:
\begin{equation}
\xi_q=\left \{
\begin{array}{ll}
     q/\alpha, \quad & (0<q<\alpha) \\
     1,\quad & (q>\alpha)\\
\end{array}
\right.
\label{eq:22}
\end{equation}
which is significantly different from the scaling exponents observed in the atmospheric
turbulence, as shown in Fig.~\ref{fig:3}b.

\begin{figure}
\centering
 \noindent\includegraphics[width=16pc]{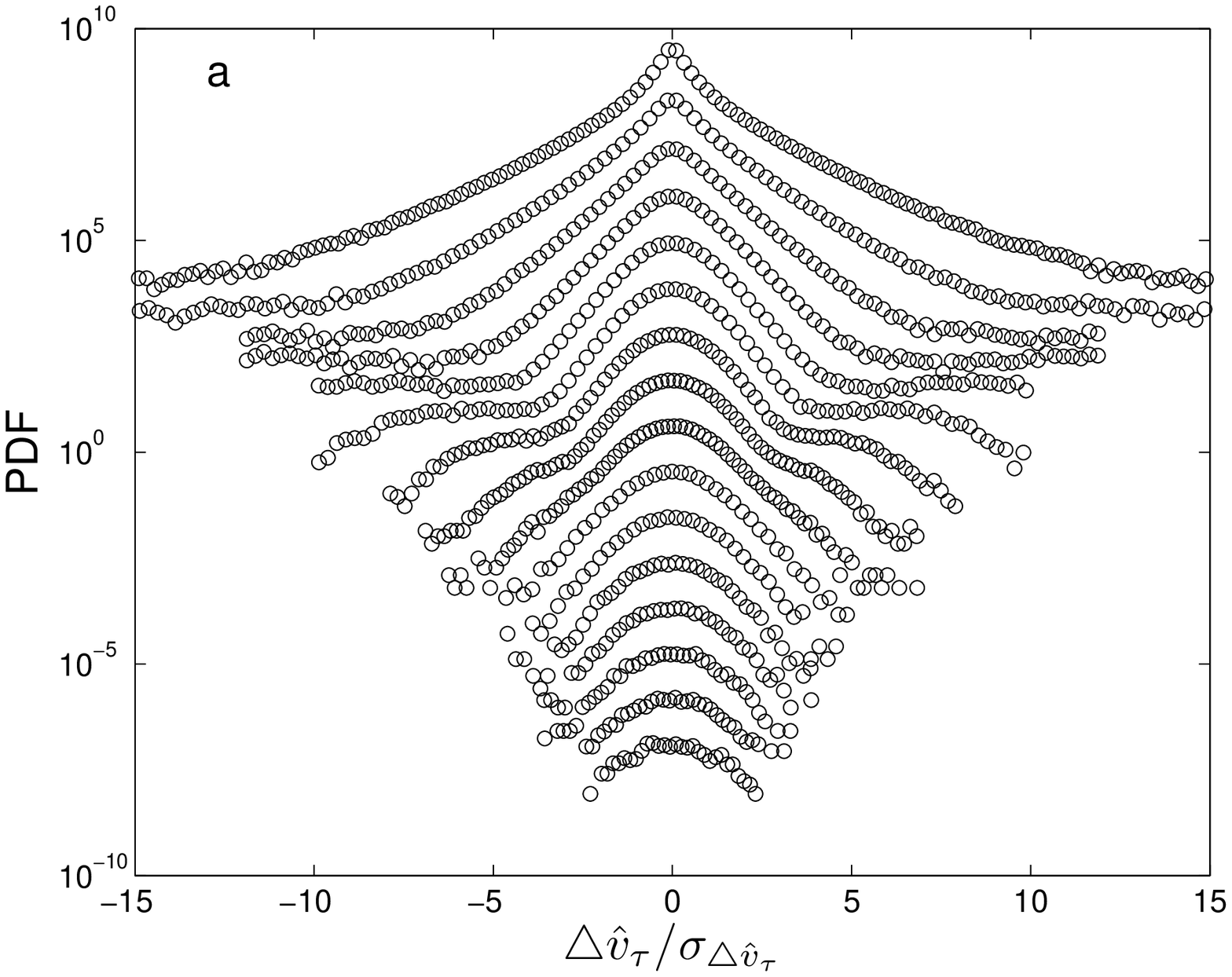}
 \noindent\includegraphics[width=16pc]{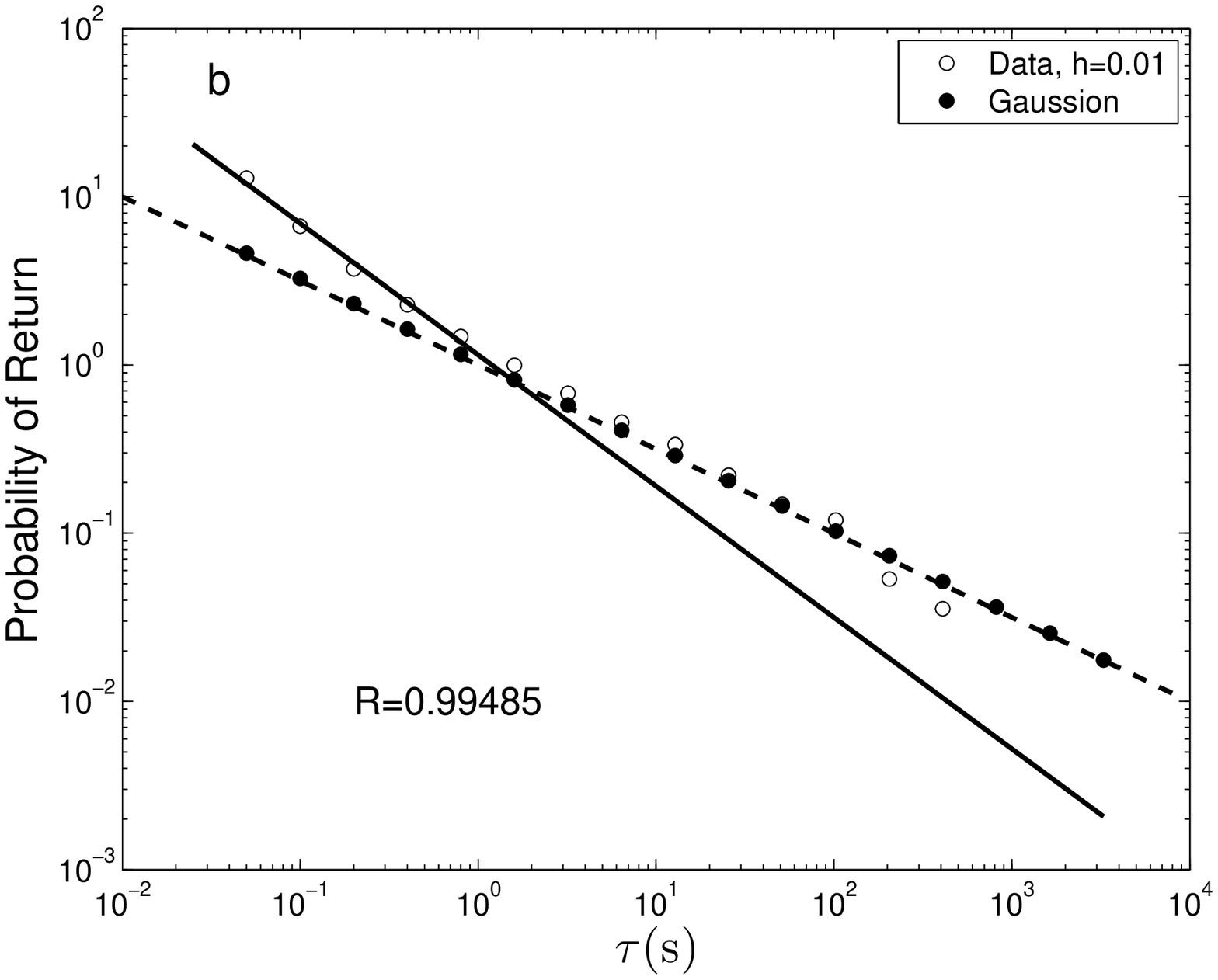}
 \noindent\includegraphics[width=16pc]{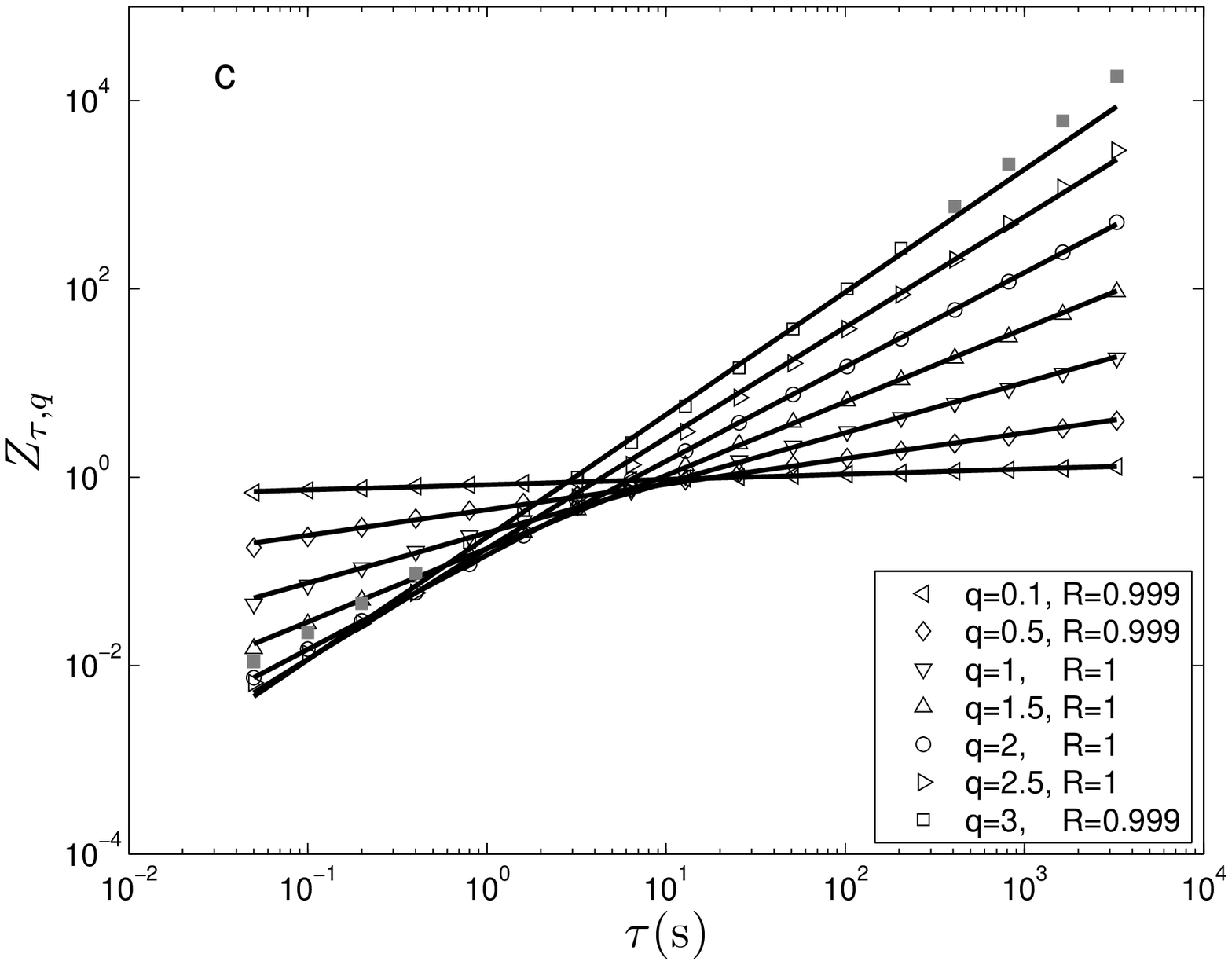}
 \noindent\includegraphics[width=16pc]{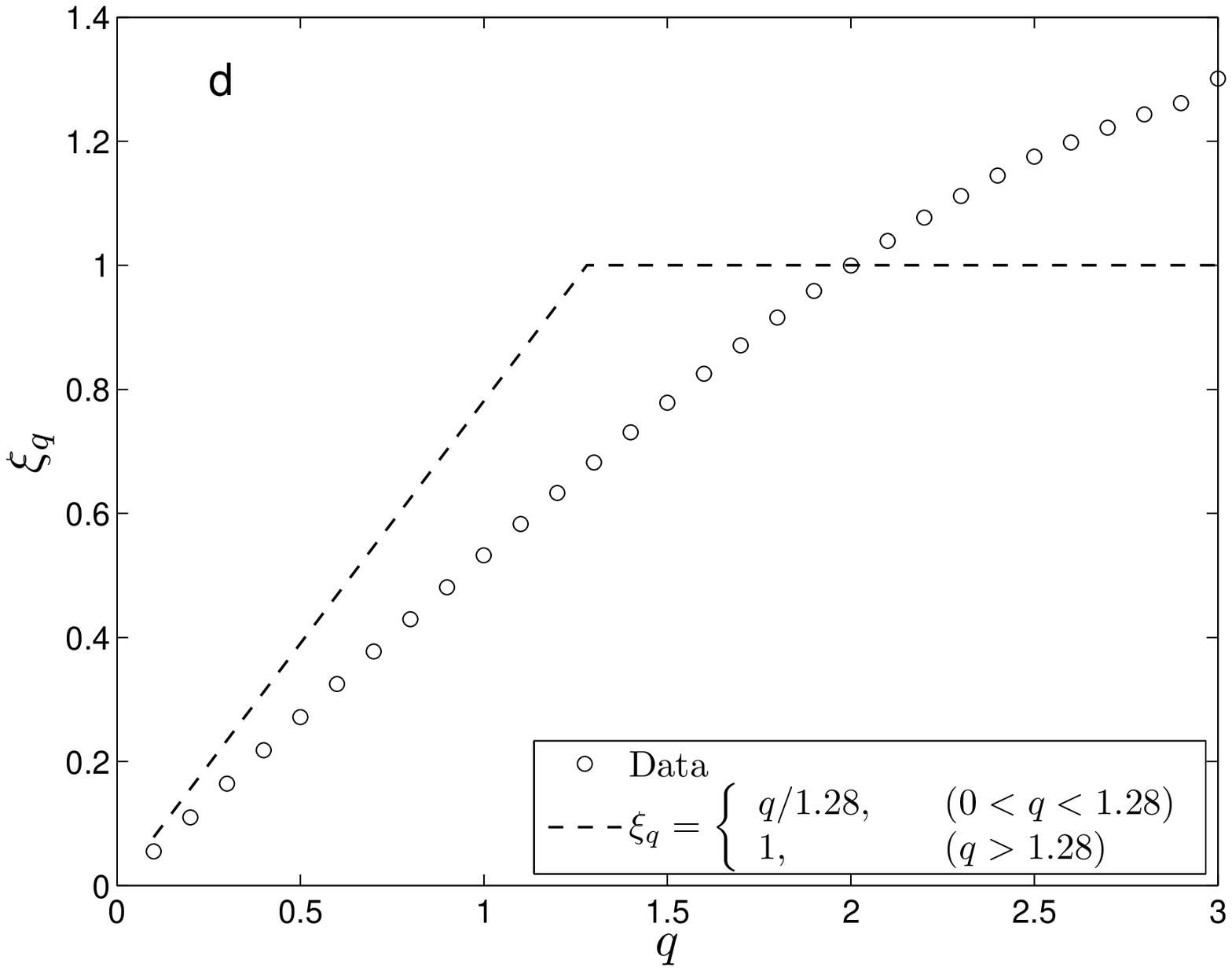}
 \caption{\label{fig:4} Results for the surrogate
 wind velocity increments without correlations: (a) Standardized PDFs
 at different time scales. From top to bottom, the corresponding
 time scales are the same as those in Fig.~\ref{fig:1}. Plots are
 also arbitrarily shifted for illustration.
 (b) Probability of return with threshold $h=0.01$
 (circles). The probability of return for Gaussian distribution
 (see Eq.~(\ref{eq:19})) is also shown here (black dots).
 The slope of the best-fit line (fitting data when $\tau<0.8$ s) is $-0.7808$ and the corresponding correlation coefficient is about $0.99$. Broken line shows a power function
 with an exponent of -0.5. (c) Scaling behavior of $q$-th order moments $Z_{\tau,q}$. All the symbols are the same as those in Fig.~\ref{fig:3}a.
 (d) Scaling exponents $\xi_q$ vary as a function of $q$. Broken lines
 shows a bi-linear function. The slope of this function is 0.7808 when $q<1/0.7808$ (see Eq.~(\ref{eq:22})).}
 \end{figure}

In order to deduce Eq.~(\ref{eq:21}), we have assumed that
wind velocity increments in
Eq.~(\ref{eq:15}) are independent of each other . However, some correlations may exist among these increments. A direct proof comes from the scaling behavior of second-order moment. As shown in Fig.~\ref{fig:3}, the scaling exponent of second-order moment is obviously smaller than one, while for
independent random variables this value equals to one. Correlation may be
a possible reason that the scaling exponents of probability of
return is smaller than the prediction of truncated stable
distribution. To verify this speculation, we generate an independent
time series by randomly sort the original time series of $\triangle
\hat{v}_{\tau_0}$. This surrogate time series will have the same
statistical distribution as the original time series. Independent surrogate time series at other larger scales can also be obtained by Eq.~(\ref{eq:3}). We then compute the
probability density functions of the surrogate time series at
different time scales and find that with the increase of scales they approach Gaussian-like
distributions much faster than the original time series
(comparing Fig.~\ref{fig:4}a with Fig.~\ref{fig:1}). This is also proved in Fig.~\ref{fig:4}b, where the probability of return is almost the same as that of independent
Gaussian distribution when time scale $\tau>1s$. Most importantly,
one can see that the scaling exponent of probability of return is
about 0.78 which falls on the range $[0.5,\infty)$ that the scaling exponent
of truncated stable distribution belongs to. Thus, we can conclude
that the correlation in the wind velocity increments
is indeed a possible reason that the scaling exponent of
probability of return deviates from the prediction of truncated
stable distribution. However, this reason is not enough to account for
the deviation of moments. For the surrogate time series,
their exponents of moments $\xi_q$ is also a concave function of $q$, while the scaling exponents of truncated stable distribution obey a bi-linear behavior.

According to Eq.~(\ref{eq:9}), the log-normal PDF model can be
expressed by a simple random mapping:
\begin{equation}
\triangle \hat{v}_{\tau} \stackrel{d}{=}
W_{\bar{\sigma}_\tau,\lambda_\tau}G\label{eq:23},
\end{equation}
where $W_{\bar{\sigma}_\tau,\lambda_\tau}$ is a log-normal random
variable and its logarithm has a mean of $\ln \bar{\sigma}_\tau$ and
a standard deviation of $\lambda_\tau$. The symbol $G$ denotes a
normally distributed random variable and is independent of
$W_{\bar{\sigma}_\tau,\lambda_\tau}$. For log-normal random
variables, we have
\begin{equation}
W_{\bar{\sigma}_\tau,\lambda_\tau} \stackrel{d}{=}
W_{\bar{\sigma}_{T},\lambda_{T}}W_{\bar{\sigma}_{\tau'},\lambda_{\tau'}},
\label{eq:24}
\end{equation}
where $W_{\bar{\sigma}_{T},\lambda_{T}}$ and
$W_{\bar{\sigma}_{\tau'}}$ are independent log-normal random
variables, and $\ln\bar{\sigma}_{T}+\ln\bar{\sigma}_{\tau'} = \ln
\bar{\sigma}_{\tau}$ and
$\lambda_{T}^2+\lambda_{\tau'}^2=\lambda_{\tau}^2$. Based on
Eq.~(\ref{eq:23}) and Eq.~(\ref{eq:24}), one can deduce that the
log-normal PDF model has a cascade process:
\begin{equation}
\triangle \hat{v}_\tau \stackrel{d}{=}
W_{\bar{\sigma}_{T},\lambda_{T}}\triangle \hat{v}_{\tau'}.
\label{eq:25}
\end{equation}
Comparing Eqs.~(\ref{eq:20}) and (\ref{eq:25}), we find that the
cascade process is very different between the truncated stable
distribution and the log-normal PDF model. For the truncated stable
distribution, the connection between different scales is a
non-random power function which can only produce a stochastic
process with scaling exponents $\xi_q$ of moments varying as a
linear function of order $q$. This process is referred as a
self-affine or self-similar fractal. To obtain a stochastic process
with a non-linear scaling exponents of $\xi_q$ which is referred as
a multifractal, one should extend the non-random power function to a
random variable \cite{mandelbrot97}. For the log-normal PDF model,
the connection is a log-normally distributed random variable and it
can produce a multifractal process by assuming suitable
relationships between parameters ($\bar{\sigma}_\tau$,
$\lambda_\tau$) and time scales $\tau$.

\begin{figure}
\centering
 \includegraphics[width=20pc]{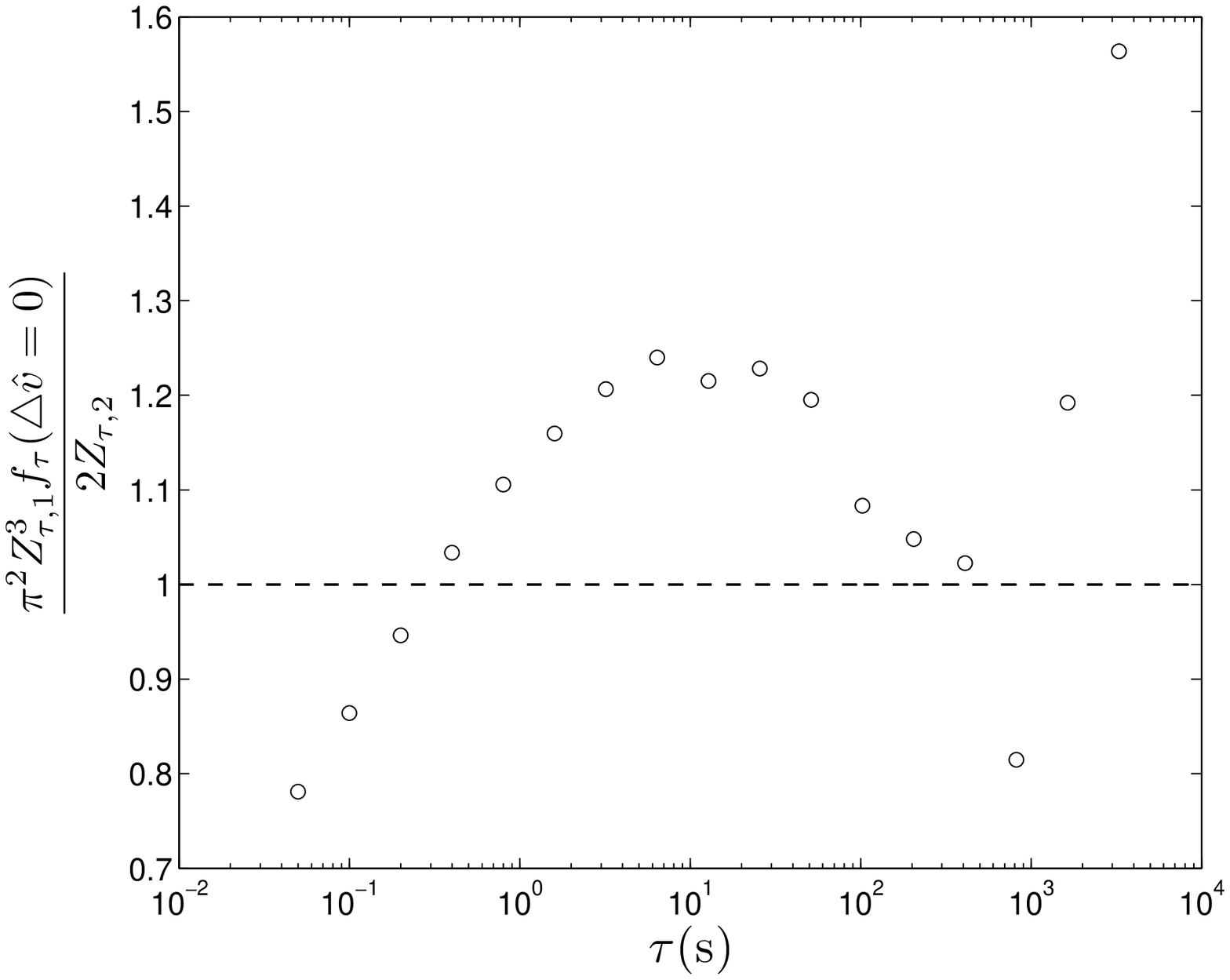}
 \caption{\label{fig:5} Test of Eq.~(\ref{eq:29}).}
 \end{figure}

As we have mentioned in Sec.~\ref{mome}, under the assumptions of
Eq.~(\ref{eq:12}) and $\bar{\sigma}_\tau\propto \tau^b$, the scaling
exponents $\xi_q$ of log-normal model vary as a parabola function
(see Eq.~(\ref{eq:14})) and this function can fit the data well (see
Fig.~\ref{fig:3}b). However, we will show that whatever the
relationships between parameters ($\bar{\sigma}_\tau$,
$\lambda_\tau$) and time scales are assumed the log-normal PDF model
can not fit the probability of return and the moments at the same
time. According to Eq.~(\ref{eq:9}), the probability of return of
log-normal PDF model is:
\begin{equation}
f_\tau(\triangle \hat{v}=0)=\frac{e^{\lambda_{\tau}^2/2}}{\sqrt{2\pi}\bar{\sigma}_\tau}
\label{eq:26}
\end{equation}
We can also deduce its $q$-order moment \cite{cgh91}:
\begin{equation}
Z_{\tau,q}={\bar{\sigma}_\tau}^qP_q\exp(q^2{\lambda_\tau}^2/2),
\label{eq:27}
\end{equation}
where
\begin{equation}
P_q=\displaystyle \frac{1}{\sqrt{2\pi}}\int_{-\infty}^{\infty} |x|^q
\exp(-x^2/2)\, d x. \label{eq:28}
\end{equation}
From Eqs.~(\ref{eq:26}) and (\ref{eq:27}), we have:
\begin{equation}
\frac{\pi^2}{2}\frac{Z_{\tau,1}^3f_\tau(\triangle
\hat{v}=0)}{Z_{\tau,2}}=1 \label{eq:29}
\end{equation}
In Fig.~\ref{fig:5}, one can see that the left-hand side of
Eq.~(\ref{eq:29}) is not a constant of one and varies as a function
of $\tau$. Thus, we conclude that the log-normal PDF model can not
fit the probability of return and the moments at the same time.

\section{Conclusions}
\label{conc}

In this paper, we have shown the scaling behavior of the probability
of return and the moments of atmospheric turbulent velocity
increments. The probability of return is found to vary as a power
function of time scale at smaller time scales. This scaling behavior
is observed for time scales spanning three orders of magnitude, from
0.1s to 100s. The moments are also found to vary as a power function
of time scales at smaller time scales. The scaling range can also
spanning three orders of magnitude when the order of moment is not
too large. However, with the increase of the order of moment the
scaling range begins to shrink. Since the behavior of higher-order
moments is mainly defined by tail distributions, the shrinkage may
be caused by different cascade processes that the central and tail
distributions represent. This conclusion should be carefully
screened by using more data. Scaling exponents of moments vary as a
concave function of order, which suggests that the time series of
atmospheric turbulent velocity increments is a multifractal. Scaling
exponents of moments observed in the homogeneous and isotropic
turbulence are also concave but the values are obviously larger than
those observed in the atmospheric turbulence.

The scaling behavior of the probability of return and the moments
observed in the time series of atmospheric turbulent velocity
increments is compared with the predictions of two commonly used PDF
model, the truncated stable distribution and the log-normal PDF
model. We find that both PDF models can not fit the probability of
return and the moments at the same time. The truncated stable
distribution can produce a power law of probability of return but
the power exponent is too large to fitting the data. One possible
reason is that the truncated stable distribution can only describe
the independent random process while some statistical correlations
always exist in the time series of atmospheric turbulent velocity
increments. The truncated stable distribution can also produce a
power law of moments. The power exponents vary as a bi-linear
function of order, while the observed exponents are concave. The
log-normal PDF model seems to be better than the truncated stable
distribution. Under some conditions, its exponents vary as a
parabola function which can fit the data well. However, this model
can not fit the probability of return and the moments at the same
time either.

Although above mentioned PDF models can not describe the scaling
behavior in the atmospheric turbulence, they also give a clues that
the scaling behavior may related to the cascade process of turbulent
eddies. Our study may help in the understanding of the cascade
mechanism in the atmospheric turbulence.

\section*{Acknowledgements}
This work is supported by the National Nature Science Foundation of
China under Grant No.~41105005.


\bibliographystyle{model1-num-names}
\bibliography{reference}

\end{document}